\documentclass[english,showpacs,preprint,superscriptaddress]{revtex4-1}

 \UseRawInputEncoding

\usepackage{babel}
\usepackage[T1]{fontenc}      
\usepackage{mathrsfs}         
\usepackage{bm}
\usepackage{color, xcolor}   
 
\usepackage{lmodern}     

\usepackage{amsmath,amsfonts,amssymb}
\usepackage{esint}     
\usepackage{extarrows}

\usepackage[ruled,linesnumbered]{algorithm2e}
\usepackage{algpseudocode}

\usepackage{graphicx}   
\usepackage{float}
\usepackage{tikz}       
\usepackage{palatino}
\usetikzlibrary{arrows,shapes,chains}

\usepackage{subfig}     
\usepackage{booktabs}  

\usepackage{makecell,rotating,multirow,diagbox}  

\usepackage[justification=centering]{caption}
\usepackage{siunitx}

\usepackage{indentfirst}
\usepackage{setspace}

\makeatletter

\usepackage[title]{appendix}

\usepackage{hyperref}  
\hypersetup{
    breaklinks = true,
    colorlinks = true,
    citecolor = {blue},
    urlcolor = {blue},
    linkcolor = {blue}
} 

\usepackage{cleveref}

 \global\long\def\calK{\mathcal{K}}
 
 \global\long\def\calM{\mathcal{M}}

 \global\long\def\calMh{\hat{\mathcal{M}}}

 \global\long\def\calRh{\hat{\mathcal{R}}}

 \global\long\def\scrf{{f}}

 \global\long\def\scrfh{\hat{{f}}}

 \global\long\def\frakCh{\hat{\mathfrak{C}}}

 \global\long\def\rmP{\mathrm{P}}

 \global\long\def\rmd{\mathrm{d}}

 \global\long\def\rmv{v}

 \global\long\def\rmvh{\hat{{v}}}

 \global\long\def\bbN{\mathbb{N}}

 \global\long\def\nh{\hat{n}}
 \global\long\def\uh{\hat{u}}

 \global\long\def\Pl{\rmP_l}

 \global\long\def\lMI{l_{M_1}}

 \global\long\def\mub{{\mu}_\beta}

 \global\long\def\wmub{w_{{\mu}_\beta}}

 \global\long\def\ddt{\frac{\partial}{\partial t}}

 \global\long\def\sumlolM{\sum_{l=0}^{l_M} }

 \global\long\def\sumbIlMI{\sum_{\beta=1}^{\lMI} }

 \global\long\def\nbvth{\frac{n_b}{\vbth^3}}

 \global\long\def\fl{f_l}

\global\long\def\fh{\hat{f}}

 \global\long\def\fhl{\fh_l}

\global\long\def\calMjl{\calM_{j,l}}

\global\long\def\calMhjl{\calMh_{j,l}}

\global\long\def\calRhjl{\calRh_{j,l}}

  \global\long\def\calRhab{{\calRh}{_{ab}}}

\global\long\def\calRhabjl{\calRhab_{j,l}}

 \global\long\def\Kl{\calK_l}
 
 \global\long\def\Klm{\calK_l^m}

 \global\long\def\scrfI{\scrf{_1}}

 \global\long\def\scrfI0{\scrf{_1^0}}

 \global\long\def\scrfhI{\scrfh{_1}}

 \global\long\def\scrfhI0{\scrfh{_1^0}}

 \global\long\def\Gabh{\hat{\Gamma}_{ab}}

 \global\long\def\colhla{{\frakCh{_l}}}

 \global\long\def\colhab{\frakCh_{ab}}

 \global\long\def\colhlab{{\frakCh{_l}}_{ab}}

\global\long\def\uzhar{\uh_{a_r}}

\global\long\def\uzharl{\uh_{a_{l,r}}}

 \global\long\def\vath{\rmv_{ath}}

 \global\long\def\vhathr{\rmvh_{{ath}_r}}

 \global\long\def\vhathrl{\rmvh_{{ath}_{l,r}}}

 \global\long\def\vbth{\rmv_{bth}}

\global\long\def\nhar{\nh_{a_r}}
\global\long\def\nharl{\nh_{a_{l,r}}}

\global\long\def\NKa{N_{K_a}}

 \global\long\def\FIG#1{~\ref{#1}}
 \global\long\def\EQ#1{~(\ref{#1})}

 \global\long\def\SEC#1{~\ref{#1}}

\makeatother

\begin{document}
 
\title{Higher-order moment convergent method in weakly anisotropic plasma and the NLVFP code for solution of the 0D-2V Vlasov-Fokker-Planck equation}

\author{Yanpeng Wang}
\email{E-mail: tangwang@mail.ustc.edu.cn}
\affiliation{School of Nuclear Sciences and Technology, University of Science and Technology of China, Hefei, 230026, China}
  

\begin{abstract}

  Fusion plasma and space plasma are typical non-equilibrium and nonlinear systems, with the interactions between different species well described by the Vlasov-Fokker-Planck (VFP) equations. The transport of mass, momentum, energy, and temperature relaxation are important issues, which are affected by the collision term of VFP even in so-called collisionless plasma domain. Hence, nonlinearity and collisions are important features in large regime. A successful numerical simulation for non-equilibrium plasma has to be able to conserve mass, momentum and energy, while satisfying Boltzmann's H-theorem and higher-order moment convergence. An expansion of the distribution function in spherical harmonics (Legendre basis when the velocity space exhibits axisymmetry) in angle coordinate and in King basis in speed coordinate of velocity space is well suited to address these requirements. This paper reviews the formulation of the 0D-2V VFP equation in terms of spherical harmonics coupled with King function and its solution in our NLVFP code. In this topic review, we will introduce the background physics related to the nonlinear VFP simulation, then describe NLVFP for 0D-2V homogeneous, weakly anisotropic plasma with utilization of the Shkarofsky's form of Fokker-Planck-Rosenbluth (FPRS) collision operator. 

  
\end{abstract}

\maketitle 

     


 \UseRawInputEncoding


\begin{spacing}{0.3}  

\section{Introduction}
\label{Introduction}

In fusion plasma relevant to conventional magnetic confinement fusion (MCF) and inertial confinement fusion (ICF), the transport processes between the main-body plasma species,
as well as the wave-particles interactions, typically play a dominant role in shaping the evolution of the non-equilibrium and nonlinear plasma system\cite{Chen1984}. Transport, since it involves collisions, has traditionally been modelled with Vlasov-Fokker-Planck (VFP) codes. 
However, simulations for these transport processes usually face challenges to conserve mass, momentum and energy\cite{Thomas2012}.
The fundamental difficulty in simulations lies in the need to include both collisionless fast particles (such as high-energy $\alpha$ particles) and highly collisional thermal particles (fusion fuels) into a consistent framework at the same time. 
While hybrid codes\cite{xia2017} separate the two distributions for treatment, they encounter issues such as fast particles escape effects, non-Maxwellian distortion of the thermal electron distribution, and thermal particles being elevated to fast particles in the absorption region of ion cyclotron resonance heating (ICRH) or electron cyclotron resonance heating (ECRH). 
That is saying for certain problems, it is imperative to treat the particle population as a unified entity rather than separating them.
Additional difficulties arise from the disparate thermal velocities\cite{Larroche2003,Taitano2016,wang2024Aconservative} due to the significant mass discrepancy ($i$-$e$ collisions) or energy difference ($i$-$\alpha$ collisions), and a linearized model is usually conventionally chosen\cite{Bell1981,Matte1982,Alouani-Bibi2005,Bell2006,Tzoufras2011}. 

An effective approach based on the finite volume method\cite{MortonK2005Numerical} (FVM) is employed to solve the 0D-2V nonlinear VFP equation\cite{Taitano2015Amass,Taitano2015II,Taitano2016,Taitano2017,Daniel2020}, incorporating the divergence form of Fokker-Planck-Rosenbluth\cite{Taitano2016} (FPRD) collision operator. 
This approach has successfully preserved mass, momentum and energy, satisfying Boltzmann's H-theorem\cite{Boltzmann1872}, while it is still a challenge to achieve the higher-order moment convergence.
Other moment convergent kinetic approaches, such as discrete Boltzmann method\cite{Zhang2019} (DBM), and lattice Boltzmann method\cite{xiao2024bohm,zhu2023} (LBM) have demonstrated the importance of higher-order moment convergence in simulating the VFP equation (Boltzmann equation in plasma physics). However, these kinetic approaches typically rely on conventional near-equilibrium assumption, which may limit their accuracy and applicability. The above limitations can be effectively addressed by NLVFP.

Building upon the previous works by Rosenbluth\cite{Rosenbluth1957}, Bell\cite{Bell2006}, and Tzoufras\cite{Tzoufras2011} in employing the spherical harmonic expansion (SHE) method (Legendre basis will be utilized for axisymmetric velocity space), we have developed a novel higher-order moment convergent method\cite{wang2024Relaxationmodel,wang2024Aconservative} for weakly anisotropic\cite{wang2025TransportI} plasma in NLVFP. Our approach utilizes the nonlinear FPRS collision operator and incorporates a King function expansion (KFE) method following SHE, by introducing the King function in Ref.\cite{wang2024Aconservative}. This methodology, encompassing both a moment approach\cite{wang2024Relaxationmodel,wang2024General} and a meshfree approach\cite{wang2024Aconservative}, is currently introduced for solving the Vlasov-Fokker-Planck (VFP) equation in scenarios where the plasma is weakly anisotropic\cite{Bell2006,wang2024Aconservative}. In contrast to conventional methods such as Grad's moment method\cite{Grad1949}, DBM\cite{Zhang2019}, and LBM\cite{zhu2023}, which are based on the near-equilibrium assumption, this moment convergent method relies on a finitely distinguishable independent features\cite{Teicher1963,Yakowitz1968} hypothesis introduced in Ref.\cite{wang2024Relaxationmodel}. 
Under this hypothesis, we have presented a series of studies\cite{wang2024Aconservative,wang2024Relaxationmodel,wang2024General}.
A general relation model called the kinetic moment-closed model\cite{wang2024Relaxationmodel,wang2024General} is derived from the VFP equation for homogeneous plasma under spherical symmetry and axisymmetry in velocity space respectively, which can be extended to the particle distribution function (PDF) with three-dimension velocity space. A conservative, implicit solver for 0D-2V multi-species nonlinear VFP equation in the meshfree approach is introduced in Ref.\cite{wang2024Aconservative}.

The key innovation of this moment convergent method lies in the introduction of new functions, specifically the King introduced in Ref.\cite{wang2024Relaxationmodel} and R function introduced in Ref.\cite{wang2024General}. 
This framework enables the achievement of a relaxation model\cite{wang2024Relaxationmodel} for homogeneous plasma and its extension to a general relaxation model\cite{wang2024General}. 
This general relaxation model allows for the derivations of a general temperature relaxation model and the explicit determination of the general characteristic frequency of temperature relaxation in the presence of spherical symmetry in velocity space. 
Furthermore, when the velocity space is axisymmetric, an optimal implicit meshfree algorithm also has been introduced for solving the coupled multi-species nonlinear VFP equation in 0D-2V\cite{wang2024Aconservative}. The key to this meshfree approach is expanding the solutions in terms of the Legendre basis (in polar angle coordinate) and King basis (in speed coordinate). The Legendre polynomial expansion converges exponentially, while the novel King method serves as a moment convergent algorithm that ensures conservation with high precision in discrete form. Both of these moment convergent methods satisfy conservation laws, Boltzmann's H-theorem and higher-order moment convergence, making them suitable for studying nonlinear evolution at multi-time and multi-space scales in weakly anisotropic plasma physics.


\end{spacing}

\begin{spacing}{0.2}   

\section{NLVFP}
\label{NLVFP}

\subsection{Expansions utilized in NLVFP}
\label{Expansions utilized in NLVFP}

NLVFP, shorts for nonlinear VFP, is currently a zero-dimensional in physics space, multi-dimensional in velocity space VFP code, employing expansion methods. In NLVFP, the 0D-2V normalized PDF is represented in normalized velocity space ($\rmvh,\mu$) by an expansion in Legendre basic\cite{Arfken1971}, $\Pl $ and King basis\cite{wang2024Aconservative}, $\Kl$:
  \begin{eqnarray}
      \fh \left(\rmvh,\mu,t \right) &=& \sumlolM \fhl \left( \rmvh,t \right) \Pl \left(\mu \right), \quad 0 \le l \le l_M, \label{fhfhl}
  \end{eqnarray}
where
    \begin{eqnarray}
        \fhl \left(\rmvh,t \right) &=& C_3 \sum_{r=1}^{\NKa}  \left [ \nharl \Kl \left(\rmvh;\uzharl,\vhathrl \right) \right] ~.  \label{KFE}
    \end{eqnarray}
Here, $C_3=\sqrt{2 \pi} / {\pi^{3/2}}$, $l \in \bbN$, $l_M \in \bbN$, $r \in \bbN^+$, $\NKa \in \bbN^+$. $n_a$ and $\vath$ are respectively the number density and thermal velocity of species $a$. Normalized speed of species $a$, $\rmvh = v / \vath$ and normalized PDF of species $a$, $\fh =  n_a^{-1} \vath^3 f$. 
Parameters $\uzhar$ and $\vhathr$ are the characteristic parameters of the $r^{th}$ King function with weight function $\nhar$. 
The convergence of King function expansion (KFE) is established by utilizing Wiener's Tauberian theorem\cite{Wiener1932} and demonstrated in Ref.\cite{wang2024Relaxationmodel} under the finitely distinguishable independent features\cite{Teicher1963,Yakowitz1968} (FDIF) hypothesis. This hypothesis\cite{wang2024Relaxationmodel} indicates that $\NKa$ will typically be a finite number, usually no more than 10 for a weakly anisotropic plasma. The definition of anisotropy is provided in Ref.\cite{wang2025TransportI}.

Eq.\EQ{KFE}, where the characteristic parameters are typically dependent on $l$, represents the general King mixture model (GKMM) when velocity space exhibits general axisymmetry, and the responding plasma state referred to as local sub-equilibrium. If the characteristic parameters in KFE represented by Eq.\EQ{KFE} are independent on $l$, GKMM reduces to as the King mixture model (KMM) and the plasma state described by KMM is in a local quasi-equilibrium. Comparing the GKMM and KMM, it can be observed that KMM degenerates in angle. Furthermore, if KMM degenerates in speed ($\NKa \equiv 1$), it reduces to be the King model (KM), with the responding state referred to as local-equilibrium. However, if KMM further degenerates in angle ($\fhl \equiv 0$ when $l \ge 1$), it reduces to be the zeroth-order King mixture model (KMM0). In this situation, the velocity space exhibits general spherical symmetry and the responding state will be referred to as quasi-equilibrium. If there is no shell structure present in the spherical symmetric velocity space, which can be described by Maxwellian mixture model (MMM), then the responding state will be denoted as shell-less quasi-equilibrium. Furthermore, if MMM degenerates in speed ($\NKa \equiv 1$) when velocity space exhibits general spherical symmetry, it reduces to be Maxwellian model (MM) and the corresponding plasmas state is thermodynamic equilibrium. For isotropic, weakly anisotropic and moderately anisotropic systems\cite{wang2025TransportI}, the plasma states are presented in Table\FIG{tab:States} based on the symmetry of velocity space and the corresponding King-type models. For the scenario with general velocity space, the corresponding descriptive models is currently under development and will be presented in the future.

  \begin{table}[H]
  \captionsetup{font=footnotesize}
      \renewcommand\arraystretch{1.3}
      \centering
      \caption{States of plasma: (isotropic, weakly anisotropic and moderately anisotropic system) according to the symmetry of velocity space and the corresponding King-type models.}
      \label{tab:States}
      \begin{tabular}{c|c|c|c}
      \hline
          \multicolumn{2}{c|}{Symmetry of velocity space} & Plasma states & Models
          \\ \hline 
          \multirow{3}{*}{\makecell{Spherical symmetry \\ (isotropic)}} & Degenerates in speed & Thermodynamic equilibrium & MM
          \\  \cline{2-4}
                                         & Without shell structure & Shell-less quasi-equilibrium & MMM
          \\  \cline{2-4}
                                         & General & Quasi-equilibrium & KMM0
          \\ \hline
          \multirow{3}{*}{\makecell{Axisymmetry \\ (weakly anisotropic and \\ moderately anisotropic)}}  & Degenerates in speed & Local-equilibrium & KM
          \\  \cline{2-4}
                                        & Degenerates in angle  & Local quasi-equilibrium & KMM
          \\  \cline{2-4}
                                        & General & Local sub-equilibrium & GKMM
          \\ \hline
          General velocity space & \multicolumn{3}{c}{In developing}
          \\ \hline
      \end{tabular}
  \end{table}

  \begin{figure}[H]
	\begin{center}
		\includegraphics[width=0.7\linewidth]{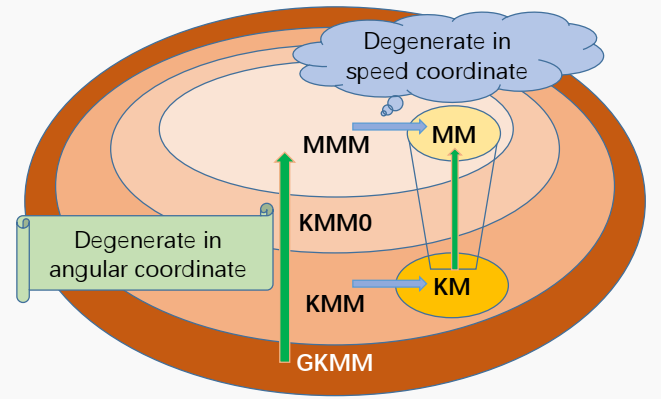}
	\end{center}
	\caption{Illustrating scopes of King-type models: Various levels based on degeneracy in velocity space.}
	\label{StatesGKMM}
  \end{figure}

The relationships between different King-type models are illustrated in Fig.\FIG{StatesGKMM}. It is apparent that MM, MMM, KMM0, KM, and KMM are all specialized models of GKMM. Furthermore, GKMM, KMM, KMM0 and MMM exhibit progressive degeneracy in angular coordinate, similar to the relationship between KM and MM. Additionally, varying levels of GKMM, KMM0, MMM with various $\NKa$ demonstrate progressive degeneracy in speed coordinate; a similar trend is observed for the relationship between KMM and KM as well as between MMM and MM. It is noteworthy that MM, MMM and KMM0 are isotropic models. Meanwhile, both KMM and GKMM can be classified as weakly anisotropic\cite{wang2025TransportI} models when all characteristic parameters satisfy $\uzhar/\vhathr \ll 1$ and moderately anisotropic\cite{wang2025TransportI} models when $\uzhar/\vhathr \le 3$ for all $r^{th}$ sub-component in KFE\EQ{KFE}. Our presented works\cite{wang2024Relaxationmodel,wang2024General,wang2024Aconservative} are focusing on the weakly anisotropic plasma system.

Bell et al.\cite{Bell2006} have  presented a comprehensive overview of the benefits and drawbacks of utilizing SHE in angular coordinate of velocity space coupled with a finite difference method (FDM) in speed coordinate. In addition to the two limitations mentioned by Bell, such as negative of amplitude of distribution function and challenge near the point where $\rmvh =0$, momentum conservation and fully nonlinearity also pose challenges for their algorithm. However, all these obstacles can be overcome by employing the KFE approach.

\subsubsection{Some advantages of KFE}
\label{Some advantages of KFE}

\begin{spacing}{0.2}  
\begin{itemize}
    \item[(i)] NLVFP will not be limited by the near-equilibrium assumption, commonly utilized in conventional methods such as LBM/DBM and Grad's moment approach. This is because KFE is based on the FDIF hypothesis\cite{wang2024Relaxationmodel}.
    \item[(ii)] King function serves as a one-dimensional shape function with continuous smoothness, which also acts as the smoothing step for calculating the PDF.
    \item[(iii)] KFE successfully navigates the challenge near the point where $\rmvh =0$ in SHE, due to the smoothness of the King function.
    \item[(iv)] KFE is able to automatically maintain the non-negativity by determining the appropriate truncation for SHE, leveraging the analytical properties of KFE.
    \item[(v)] Coulomb collisions not only make amplitudes of PDF decay exponentially in SHE due to angular scattering\cite{Bell2006}, but also lead to rapid convergence for the amplitudes in speed coordinate. KFE is a moment convergent technique that has been demonstrated to achieve up to order 16\cite{wang2024Aconservative}. Therefore, conservation can be achieved with high precision in a discrete manner, and high-order moment convergence can be easily addressed for any order without special treatment.
    \item[(vi)] Coupling KFE with SHE ensures the symmetry of the collision operator in discrete and accurately satisfies conservation through the application of manifold technique.
    \item[(vii)] NLVFP is able to accurately capture the full nonlinear effects of the VFP equation by utilizing the nonlinear FPRS collision operator, attributed to the rapid convergence of SHE and KFE.
    \item[(viii)] NLVFP, employing SHE and KFE, can naturally capture the isotropic Maxwellian state in velocity space, and the resulting solution exhibits robustness.
\end{itemize}
\end{spacing}

\subsubsection{Some disadvantages of KFE}
\label{Some disadvantages of KFE}

\begin{itemize}
    \item[(i)] NLVFP retains the full nonlinearity in both the angular and speed coordinate of velocity space, resulting in a certain degree of increase in computational complexity.
    \item[(ii)] The characteristic parameters in KFE should be determined using alternative methods\cite{wang2024Relaxationmodel,wang2024Aconservative}, such as solving the characteristic parameter equations as described in Sec.\SEC{Characteristic parameter equation}.
    \item[(iii)] The current NLVFP is limited to weakly anisotropic plasma due to the assumption that characteristic parameters in KFE are independent of $l$, resulting in the code's inability to handle moderately anisotropic scenarios where the average velocity is comparable to thermal velocity.
\end{itemize}
\noindent
The initial two challenges, (i) and (ii), are inherent outcomes of the nonlinearity and have not been observed to pose significant issues individually. The third limitation can be addressed by expanding to a more general version for moderately anisotropic plasma, where the characteristic parameters are dependent on $l$. This enhancement will be further developed in future studies.

Expansion in spherical harmonics and King function might appear complicated, but in fact the equations turn out to be fairly straightforward and robust due to the following three reasons: (A) Many natural statistical systems, including plasma system, are independent and identically distributed systems that often adhere to the central limit theorem \cite{Rosenblatt1956}, allowing their probability density functions to be described utilizing Gaussian functions. (B) Spherical harmonics are natural bases for angular calculations in 3D, as found in many branches of physics. (C) King function is the form of Gaussian function in spherical coordinate system, which can be obtained by employing SHE.

\subsection{The equations solved by NLVFP}
\label{The equations solved by NLVFP}

NLVFP currently solves the VFP equation in zero spatial and two velocity dimensions. This framework can be extended into three velocity dimensions by introducing a more general function named as associate King function, $\Klm$, and three spatial dimensions by including the Vlasov parts\cite{wang2025TransportI}, which will be discussed in the future. 

\subsubsection{Meshfree approach}
\label{Meshfree approach}

The 0D-2V NLVFP equations in the meshfree approach\cite{wang2024Aconservative} are VFP spectrum equation, which can be written in the form
  \begin{eqnarray}
      \ddt  \fl \left(\rmvh,t \right) &=& \frac{n_a}{\vath^3} \colhla, \quad 0 \le l \le l_M \label{VFPl}
  \end{eqnarray}
for the evolution of all amplitudes of distribution function, $\fl = n_a \vath^{-3} \fhl$, with maximum order of amplitudes $l_M$. In contrast to traditional usage of the simplified model of FPRS collision operator\cite{Bell2006,Bell1981,Matte1982,Shkarofsky1992,Alouani-Bibi2005,Tzoufras2011}, the $l^{th}$-order normalized amplitude of collision operator utilized in NLVFP is the full nonlinear form of FPRS, which is described as:
  \begin{eqnarray}
      \colhla \left(\rmvh,t \right) &=& \sum_b  \nbvth \Gabh \colhlab ~. \label{colhla}
  \end{eqnarray}
Function $\colhlab$ represents the $l^{th}$-order amplitude of normalized FPRS collision operator for species $a$ colliding with species $b$, as given in Ref.\cite{wang2024Aconservative}, and can be quoted as:
  \begin{eqnarray}
      \colhlab \left(\rmvh,t \right) = \ & \sumbIlMI \wmub \Pl (\mub) \colhab \left (\rmvh, \mub, t \right)
      ~. \label{FPRS0D1V}
  \end{eqnarray}
 Here, $\lMI = l_M + 1$, $\wmub$ is the $\beta^{th}$ weight of Gauss-Legendre quadrature. Function $\colhab$ represents the normalized FPRS collision operator, which can be calculated directly based on the values of $\fhl$.  It has been demonstrated in Ref. \cite{wang2024Aconservative} that this meshfree method, utilizing SHE and KFE techniques, exhibits moment convergence.

\subsubsection{Moment approach}
\label{Moment approach}

The 0D-2V NLVFP equations in the moment approach is the kinetic moment evolution equation, serving as a transport equation and can be written in the form
\begin{eqnarray}
    \ddt \calMjl \left(t \right) &=& \rho_a \left(\vath \right)^{j} \calRhjl, \quad 0 \le l \le l_M , \quad j \ge - 2 - l \label{dtMhjl}
\end{eqnarray}
for the evolution of all kinetic moments, $\calMjl$.  
Here, the $(j,l)^{th}$-order kinetic moment and normalized kinetic dissipative force of species $a$ are respectively defined as:
\begin{eqnarray}
    \calMjl \left(t \right) & = & \rho_a \left(\vath \right)^{j} \calMhjl = 4 \pi \rho_a (\vath)^j \int_0^{\infty} \rmvh^{j+2} \fhl \rmd \rmvh, \quad j \ge - 2 - l, \label{Mjl0D2V}
    \\
    \calRhjl \left(t \right) & = & \sum_b \nbvth \Gabh \calRhabjl, \quad j \ge - 2 - l, \label{Rjl0D2V}
\end{eqnarray}
where mass density $\rho_a = m_a n_a$ and the normalized kinetic dissipative force between species $a$ and species $b$
\begin{eqnarray}
    \calRhabjl \left(t \right)  =  4 \pi 
    \int_0^{\infty} \rmvh^{j+2} \colhlab \rmd \rmvh , \quad j \ge -2 ~. \label{Rhabjl0D2V}
\end{eqnarray}

The kinetic dissipative force closure relation, as represented by Eq.\EQ{Rjl0D2V}, is solely dependent on the mass ratio, thermal velocity ratio and characteristic parameters of King function. Its analytical expression has been derived for shell-less spherical symmetric velocity space\cite{wang2024Relaxationmodel} under the FDIF hypothesis. This expression is then extended to general spherical symmetric velocity space through the introduction of the R function\cite{wang2024General}. These results can also be extended to scenarios where the velocity space exhibits axisymmetry and will be published in the future.

The combination of the transport equation\EQ{dtMhjl}, kinetic dissipative force closure relation\EQ{Rjl0D2V}, analytical expression of $\calRhabjl$ and the characteristic parameter equations\EQ{CPEs} constitutes a set of nonlinear equations, which is referred to as kinetic moment-closed model (KMCM)\cite{wang2024Relaxationmodel, wang2024General}. 
A general temperature relaxation model\cite{wang2024General} for spherical symmetric velocity space has been derived from this general relaxation model. Furthermore, Ref.\cite{wang2024Relaxationmodel, wang2024General} has demonstrated that the classic models such as two-temperature thermal equilibrium model, zeroth-order Braginskii heat transfer model and thermodynamic equilibrium model are special cases of this general relaxation model.

\subsubsection{Characteristic parameter equation}
\label{Characteristic parameter equation}

Under the FDIF hypothesis, the characteristic parameters $\nhar$, $\uzhar$ and $\vhathr$ in Eq.\EQ{KFE} and kinetic dissipative force closure relation\EQ{Rhabjl0D2V} are solely functions of any $3\NKa$ different orders of normalized kinetic moments. Those parameters can be determined by solving the characteristic parameter equations (CPEs)
      \begin{eqnarray}
            \calMhjl \left(t \right) &=& {C_M}_j^l \sum_{r=1}^{\NKa} \nhar (\vhathr)^j \left(\frac{\uzhar}{\vhathr} \right)^l 
            \left [1 + \sum_{k=1}^{(j-l)/2} C_{j,l}^k \left (\frac{\uzhar}{\vhathr} \right)^{2k} \right] , \label{CPEs}
      \end{eqnarray}
where $(j,l) \in \left\{(l+2j_p-2,l)|j_p \in \bbN^+, 0 \le l \le l_M \right\}$.
Here, the coefficients ${C_M}_j^l$ and $C_{j,l}^k$ are contingent upon the parameters $k$, $j$, and $l$ as delineated in Ref.\cite{wang2024Aconservative}.
The CPEs are a system of nonlinear algebraic equations, typically served as constraint equations for the VFP spectrum equation\EQ{VFPl} or the kinetic moment evolution equation\EQ{dtMhjl}.  In Ref.\cite{wang2024Aconservative}, several algorithms are presented for solving CPEs to determine characteristic parameters, and more advanced algorithms are currently being developed.

\subsection{NLVFP: further extension and applications}
\label{NLVFP: further extension}

The aforementioned equations lie at the heart of NLVFP, a framework that utilizes the expansion method based on the FDIF hypothesis. Generally, the moment approach, resembling an analytical method that does not require field grids of velocity space, proves to be more effective than the meshfree approach which is a semi-analytical and semi-numerical method in velocity space. Both approaches in NLVFP can be extended to general scenarios, involving axisymmetric velocity space where characteristic parameters may dependent on $l$ and a general velocity space. Furthermore, we aim to develop the nD-mV NLVFP code by incorporating the Vlasov parts in our future work.

One strength of expansion method over a particle code is that it produces results without noise and provides a solid platform for the inclusion of
further physics, and over a FDM/FVM code is the ease of handling full nonlinearity and achieving higher-order moment convergence.
NLVFP has the potential for application to a wide range of plasma types, particularly in scenarios where magnetic field-dominated transport is involved. Its expansion in spherical harmonics makes it well-suited for representing the rotation of particle trajectories with algebraic expressions, without involving differentials in velocity space. 
This method is especially effective for magnetized particle transport when the distribution is weakly anisotropic, requiring only a small numbers of spherical harmonics and King functions. It is worth to note that expansion methods avoid extensive computer time spent on modeling the isotropic part of the distribution, unlike traditional particle codes. Therefore, NLVFP may find utility in MCF as well as in space and astrophysical plasmas where magnetic fields play a dominant role.

\section{Conclusion}
\label{Conclusion}

An expansion framework has been developed in NLVFP code, based on SHE and KFE methods, to solve the 0D-2V nonlinear VFP equation for weakly anisotropic plasma. This framework operates under the FDIF hypothesis rather than the conventional near-equilibrium assumption, providing significant advantages over traditional methods. 
It offers a comprehensive description of plasma states based on the symmetry of velocity space, corresponding to various levels of King-type models including GKMM, KMM, KM, KMM0, MMM and MM.
Additionally, a general relation model, known as the kinetic moment-closed model, is presented for homogeneous plasma under spherical symmetry and axisymmetry in velocity space respectively. This provides a general temperature relaxation model and explicitly determines the explicit determination of the general characteristic frequency of temperature relaxation in the presence of spherical symmetry in velocity space.

Those models can be extended to PDF with tree-dimension velocity space. 
Both approaches in NLVFP, including the moment approach and meshfree approach, adhere to conservation laws, Boltzmann's H-theorem, and higher-order moment convergence, making them suitable for studying the nonlinear evolution of weakly anisotropic non-equilibrium plasma.

\end{spacing}

\bibliographystyle{apsrev4-1}
\bibliography{Plasma}

\end{document}